\documentclass[12pt]{article}
\usepackage{epsfig,amsfonts,amssymb}
\usepackage{hyperref}

\usepackage{cite}
\usepackage{cite}

\usepackage{comment}

\def\ZZZ{{\hbox{ Z\kern-1.6mm Z}}}
\def\RRR{{\hbox{ R\kern-2.4mm R}}}
\def\CCC{{\hbox{ C\kern-2.0mm C}}}
\def\zzz{{\hbox{z\kern-1mm z}}}

\newcommand{\qeq}{{\hbox{=\kern-2.3mm ? \kern.5mm }}}
\renewcommand{\qeq}{=}

\newcommand{\eps}{\epsilon}

\newcommand{\MM}{{\cal M}}

\newcommand{\OO}{{\cal O}}

\newcommand{\wt}{\widetilde}
\newcommand{\wh}{\widehat}

\newcommand{\be}{\begin{equation}}
\newcommand{\ee}{\end{equation}}
\newcommand{\ben}{\begin{eqnarray}\displaystyle}
\newcommand{\een}{\end{eqnarray}}

\newcommand{\refb}[1]{(\ref{#1})}

\newcommand{\sectiono}[1]{\section{#1}\setcounter{equation}{0}}

\def\one{{\hbox{ 1\kern-.8mm l}}}
\def\zero{{\hbox{ 0\kern-1.5mm 0}}}

\newcommand{\bea}[1]{\begin{eqnarray}\label{#1} }
\newcommand{\eea}{\end{eqnarray}}

\newcommand{\eqref}{\refb}

\newcommand{\non}{\nonumber}

\usepackage{bm}
\usepackage[table]{xcolor}

\def\wh{\hat}
\def\wt{\tilde}

\begin{document}

\baselineskip 24pt

\begin{center}

{\Large \bf Classical Soft Graviton Theorem Rewritten}


\end{center}

\vskip .6cm
\medskip

\vspace*{4.0ex}

\baselineskip=18pt

\centerline{\large \rm Biswajit Sahoo$^a$ and  Ashoke Sen$^b$
}

\vspace*{4.0ex}

\centerline{\large \it $^a$Fields and Strings Laboratory, Institute of Physics}
\centerline{\large \it Ecole Polytechnique Federale de Lausanne (EPFL)}  
\centerline{\large \it CH-1015 Lausanne,
Switzerland}
\centerline{\large \it $^b$Harish-Chandra Research Institute, HBNI}
\centerline{\large \it  Chhatnag Road, Jhusi,
Allahabad 211019, India}


\vspace*{1.0ex}
\centerline{\small E-mail:   biswajit.sahoo@epfl.ch, sen@hri.res.in
}

\vspace*{5.0ex}

\centerline{\bf Abstract} \bigskip

Classical soft graviton 
theorem gives the gravitational wave-form at future null infinity at late retarded time
$u$ for a general classical scattering. The large $u$ expansion has three known 
universal terms: the constant term,
the term proportional to $1/u$ and the term proportional to $\ln u/u^2$, whose coefficients are
determined solely in terms of the momenta of
incoming and the outgoing hard particles, including the 
momenta carried by outgoing
gravitational and electromagnetic radiation produced during scattering. For 
the constant term, also known
as the memory effect, the dependence on the momenta carried away by the final state radiation 
/ massless particles is known as non-linear memory or null memory. It was shown earlier that for
the coefficient of the $1/u$ term the dependence on the momenta of the
final state massless particles / radiation cancels and the result can be written solely in terms of the
momenta of the incoming particles / radiation and the final state massive particles. In this note we
show that the same result holds for the coefficient of the $\ln u/u^2$ term. Our result
implies that for scattering of massless particles the coefficients of the $1/u$ and $\ln u/u^2$ terms
are determined solely by the incoming momenta, even if the particles coalesce to form a black hole
and massless radiation. We use our result to compute the low frequency flux of gravitational radiation
from the collision of massless particles at large impact parameter.

\vfill \eject

\tableofcontents

\sectiono{Introduction and summary} \label{s1}

Let us consider a general classical scattering process in which
a set of $m$ objects carrying four 
momenta $p_1',\cdots, p_m'$ in the asymptotic past come
together, interact and then disperse as a set of $n$ objects carrying four 
momenta $p_1,\cdots p_n$. We shall choose the origin of our space-time coordinate 
system close to the region where the particles interact and consider a gravitational wave
detector far away from the scattering region, whose space-time coordinates will be 
denoted by $(t,\vec x)$. Our object of interest will be the gravitational wave-form at the
detector:
\be
h_{\mu\nu}(t, \vec x) =  {1\over 2}(g_{\mu\nu}-\eta_{\mu\nu})\, ,
\ee
but we shall find it convenient to state the result in terms of a slightly different quantity
that carries the same information:
\be
e_{\mu\nu} \equiv h_{\mu\nu} -
{1\over 2} \eta_{\mu\nu} h_\rho^{~\rho} \qquad \Leftrightarrow \qquad
h_{\mu\nu} \equiv e_{\mu\nu} -
{1\over 2} \eta_{\mu\nu} e_\rho^{~\rho}\, .
\ee
All indices are raised and lowered by the flat metric $\eta_{\rho\sigma}$.
We also define 
\be 
R \equiv |\vec x|, \qquad \hat n ={\vec x\over R}, \qquad n = (1, \hat n)\, ,
\ee
and the retarded time at the detector:
\be 
u = t - t_0, \qquad t_0 = {R\over c} + \hbox{correction}\, ,
\ee
where $t_0$ is taken to be the time around which the peak of the gravitational wave-form reaches
the detector. The `correction' proportional to $\ln R$ represents the effect of the time delay due to
gravitational
drag on the gravitational waves due to the objects involved in the scattering. 
We denote by $G$ and $c$  respectively the Newton's gravitational constant and the speed of
light in flat space-time. We also use mostly + signature metric and compute
the inner products with flat metric unless mentioned otherwise. Since we are displaying
explicit factors of $c$, the zeroth component of the momentum will be given by the energy
divided by $c$. In this convention,
results based on soft theorem determine the form of $e_{\mu\nu}$ at the 
detector, up to
gauge transformation,
at late and early retarded time\cite{1806.01872,1808.03288,1912.06413,2008.04376}:
\ben \label{e6.1}
e_{\mu\nu} &=& A_{\mu\nu} + {1\over u} \, B_{\mu\nu} +
u^{-2} \, \ln\, |u|\, F_{\mu\nu} + \OO(u^{-2})+\OO(R^{-2}), \quad \hbox{for large positive u},\nonumber \\ 
&=&  {1\over u} \, C_{\mu\nu} + u^{-2} \, \ln\, |u| \, G_{\mu\nu} + \OO(u^{-2})+\OO(R^{-2}), 
\quad \hbox{for large 
negative u},
\een
where,
\ben \label{e6.2}
A^{\mu\nu} &=& {2\, G\over R\, c^3} \, \left[-\sum_{i=1}^{n}  
p_{i}^\mu \, p_{i}^\nu\, 
{1\over n.p_{i}} + \sum_{i=1}^{m}  
p_{i}^{\prime\mu} \, p_{i}^{\prime\nu}\, 
{1\over n.p_{i}'}  \right], 
\een
\ben \label{e6.2a}
B^{\mu\nu} &=& -\, {4\, G^2\over R\, c^7} \left[
\sum_{i=1}^n \sum_{j=1\atop j\ne i}^n 
{ p_{i}.p_{j}\over 
\{(p_{i}.p_{j})^2 
-p_{i}^2 p_{j}^2 \}^{3/2}} \, 
\left\{{3\over 2} p_{i}^2 p_{j}^2 - (p_{i}.p_{j})^2\right\} \right. 
\left. 
 \, {p_{i}^\mu \over n.p_{i}}\,
 (n.p_{j}\, p_{i}^\nu - n.p_{i}\, p_{j}^\nu ) \right.
\non\\ &&  \left.  - \left\{
 \sum_{j=1}^n\, p_{j}.n  \sum_{i=1}^{n} \, {1\over p_{i}.n} \, p_{i}^\mu p_{i}^\nu
 -  \sum_{j=1}^m\, p_{j}'.n \sum_{i=1}^{m} \, {1\over p_{i}'.n} \, p_{i}^{\prime\mu} p_{i}^{\prime \nu}\right\}
 \right]\, , 
 \een
 \ben \label{e6.2b}
C^{\mu\nu} &=& {4\, G^2\over R\, c^7} \Bigg[ 
\sum_{i=1}^m \sum_{j=1\atop j\ne i}^m 
{ p_{i}'.p_{j}'\over 
\{(p_{i}'.p_{j}')^2 
-p_{i}^{\prime 2} p_{j}^{\prime 2}\}^{3/2}} \, 
\left\{{3\over 2} p_{i}^{\prime 2} p_{j}^{\prime 2}  - (p_{i}'.p_{j}')^2\right\}
 {p_{i}^{\prime \mu} \over n.p'_{i}}\,
 (n.p_{j}^{\prime} \, p_{i}^{\prime\nu} - n. p_{i}^{\prime}\, p_{j}^{\prime\nu} )
 \Bigg] \, , \non\\
 \een
 \ben \label{eaddcon4}
  F^{\mu\nu} &  =&   {2\, G^3\over R\, c^{11}}\, \Bigg[ 4\,
\sum_{j=1}^n p_j.n \sum_{\ell=1}^n p_{\ell}.n 
\sum_{i=1}^{n} {p_{i}^\mu p_{i}^\nu\over p_{i}.n}  - 4\, 
\sum_{j=1}^{m} p'_j.n \sum_{\ell=1}^{m} p'_{\ell}.n \sum_{i=1}^{m} {p_{i}^{\prime\mu} p_{i}^{\prime\nu}
\over p'_i.n}
\non\\ && \hskip -.5in  
+ 4\, \sum_{\ell=1}^n p_{\ell}.n \sum_{i=1}^{n} \sum_{j=1\atop j\ne i}^{n}
 {1\over p_{i}.n} {p_{i}.p_j\over \{(p_{i}.p_j)^2 - p_{i}^2 p_j^2\}^{3/2}} 
 \{2 (p_{i}.p_j)^2 - 3 p_{i}^2 p_j^2\} \{ n.p_j \, p_{i}^\mu \, p_{i}^\nu - n.p_{i} \, p_{i}^\mu \, p_j^\nu\}
 \non\\ &&   \hskip -.5in
+2\, \sum_{\ell=1}^m p'_{\ell}.n \sum_{i=1}^{m} \sum_{j=1\atop j\ne i}^{m}
 {1\over p'_{i}.n} {p'_{i}.p'_j\over \{(p'_{i}.p'_j)^2 - p_{i}^{\prime 2} p_j^{\prime 2}\}^{3/2}} 
 \{2 (p_{i}'.p_j')^2 - 3 p_{i}^{\prime 2} p_j^{\prime 2}\} \{ n.p'_j \, p_{i}^{\prime\mu} \, p_{i}^{\prime \nu} - 
 n.p'_{i} \, p_{i}^{\prime \mu} \, p_j^{\prime \nu}\} \non\\ &&   \hskip -.5in
+  \sum_{i=1}^{n} \sum_{j=1\atop j\ne i}^{n}
\sum_{\ell=1\atop \ell\ne i}^{n}  {1\over p_{i}.n}  
{p_{i}.p_j\over \{(p_{i}.p_j)^2 - p_{i}^2 p_j^2\}^{3/2}}   
\{2 (p_{i}.p_j)^2 - 3 p_{i}^2 p_j^2\} {p_{i}.p_{\ell}\over \{(p_{i}.p_{\ell})^2 - p_{i}^2 p_{\ell}^2\}^{3/2}} \non\\ &&
\{2 (p_{i}.p_{\ell})^2 - 3 p_{i}^2 p_{\ell}^2\}  \{ n.p_j \, p_{i}^\mu  - n.p_{i} \, p_j^\mu \} \, 
\{ n.p_{\ell} \, p_{i}^\nu - n.p_{i} \, p_{\ell}^\nu \}
\Bigg]\, ,
\een
and
\ben \label{eaddcon5}
  G^{\mu\nu} &=&   - {2\, G^3\over R\, c^{11}}\, \Bigg[  2\, \sum_{\ell=1}^m p'_{\ell}.n \sum_{i=1}^{m} \sum_{j=1\atop j\ne i}^{m}
 {1\over p'_{i}.n} {p'_{i}.p'_j\over \{(p'_{i}.p'_j)^2 - p_{i}^{\prime 2} p_j^{\prime 2}\}^{3/2}} 
 \{2 (p_{i}'.p_j')^2 - 3 p_{i}^{\prime 2} p_j^{\prime 2}\}  \non\\ &&   \hskip 1in 
 \{ n.p'_j \, p_{i}^{\prime\mu} \, p_{i}^{\prime \nu} - 
 n.p'_{i} \, p_{i}^{\prime \mu} \, p_j^{\prime \nu}\} \non\\ &&   \hskip -.5in
-  \sum_{i=1}^{m} \sum_{j=1\atop j\ne i}^{m}
\sum_{\ell=1\atop \ell\ne i}^{m}  {1\over p'_{i}.n}  
{p'_{i}.p'_j\over \{(p'_{i}.p'_j)^2 - p_{i}^{\prime 2} p_j^{\prime 2}\}^{3/2}}   
\{2 (p'_{i}.p'_j)^2 - 3 p_{i}^{\prime 2} p_j^{\prime 2}\} {p'_{i}.p'_{\ell}\over \{(p'_{i}.p'_{\ell})^2 - p_{i}^{\prime 2} p_{\ell}^{\prime 2}\}^{3/2}} \non\\ &&  
\{2 (p'_{i}.p'_{\ell})^2 - 3 p_{i}^{\prime 2} p_{\ell}^{\prime 2}\}  
\{ n.p'_j \, p_{i}^{\prime\mu} - n.p'_{i} \, p_j^{\prime\mu} \} \, 
\{ n.p'_{\ell} \, p_{i}^{\prime \nu}  - n.p'_{i} \, p_{\ell}^{\prime\nu} \}
\Bigg]\, . 
\een
Note that the coefficients $A^{\mu\nu}$, $B^{\mu\nu}$, $C^{\mu\nu}$, $F^{\mu\nu}$ and 
$G^{\mu\nu}$ are given only by the momenta of the incoming and the outgoing objects and
do not depend on the details of the scattering process. One can also check that each of these
coefficients is gauge invariant, i.e.\ vanishes upon contraction with the four vector $n$.

In the above formul\ae, $A_{\mu\nu}$ represents a permanent change in metric
due to the passage of the gravitational wave, and is known as the memory effect\cite{mem1,mem2,mem3,mem4,christodoulou,thorne,bondi,1003.3486,
1401.5831,1312.6871}.
Its connection to the leading soft graviton theorem has been discussed in
\cite{1411.5745}. The coefficients $B_{\mu\nu}$ and $C_{\mu\nu}$, representing long range tail
of the gravitational wave-form, are related to the logarithmic correction to the 
subleading soft graviton theorem\cite{1912.06413}.
The coefficients $F_{\mu\nu}$ and $G_{\mu\nu}$ are related to the leading
logarithmic correction to the 
subsubleading soft graviton theorem\cite{1912.06413,2008.04376}.

If a significant fraction of energy is carried away by radiation, 
then the sum over the final state momenta in the expressions for $A_{\mu\nu}$,
$B_{\mu\nu}$ and $F_{\mu\nu}$ should include integration over outgoing flux of radiation, 
regarded as a flux of massless particles. 
$C_{\mu\nu}$ and $G_{\mu\nu}$ are 
given in terms of
incoming momenta only and are 
not sensitive to the momenta of outgoing particles or radiation.

The contribution to $A_{\mu\nu}$ due 
to the final state gravitational waves is some time referred to as 
non-linear memory\cite{christodoulou}
or null memory\cite{1312.6871}. 
This makes the computation of $A_{\mu\nu}$ somewhat hard since we need to
first find the angular distribution of the flux of energy carried away by the
gravitational waves. {\it A priori}, computation of $B_{\mu\nu}$
and $F_{\mu\nu}$
suffers from the same difficulty. However it was found in \cite{1808.03288} that
due to some miraculous cancellations, 
the contribution to $B_{\mu\nu}$ due to final state massless particles,
including gravitational waves, can be expressed in terms of the momenta of massive
objects only. 
If we denote by $\wt p_i$ the momenta carried by the final state massive objects only and
by $\tilde n$ the number of such objects,
then the modified formula takes the form:
\ben \label{ebmnmod}
B^{\mu\nu} &=& -\, {4\, G^2\over R\, c^7} \left[
\sum_{i=1}^{\tilde n} \sum_{j=1\atop j\ne i}^{\tilde n} 
{ \wt p_{i}.\wt p_{j}\over 
\{(\wt p_{i}.\wt p_{j})^2 
-\wt p_{i}^2 \wt p_{j}^2 \}^{3/2}} \, 
\left\{{3\over 2} \wt p_{i}^2 \wt p_{j}^2 - (\wt p_{i}.\wt p_{j})^2\right\} \right. 
\left. 
 \, {\wt p_{i}^\mu \over n.\wt p_{i}}\,
 (n.\wt p_{j}\, \wt p_{i}^\nu - n.\wt p_{i}\, \wt p_{j}^\nu ) \right.
\non \\ && \hskip -.5in \left.  - 
 \left\{ \sum_{j=1}^{\tilde n}\, \wt p_{j}.n \sum_{i=1}^{\tilde n} \, {1\over \wt p_{i}.n} \, 
 \wt p_{i}^\mu \wt p_{i}^\nu
 -  \sum_{j=1}^m\, p_{j}'.n \sum_{i=1}^{m} \, {1\over p_{i}'.n} \, p_{i}^{\prime\mu} 
 p_{i}^{\prime \nu}\right\}
 +
 \wt P^\mu \wt P^\nu - P^{\prime\mu} P^{\prime\nu} \right]\, ,  
 \een
 where
 \be \label{edefpp}
 P'=\sum_{i=1}^m p_i', \qquad \wt P = \sum_{i=1}^{\tilde n} \wt p_i\, .
 \ee
We emphasize that \refb{ebmnmod} is not an independent formul\ae\ but follows from 
\refb{e6.2a} after setting $p_i^2=0$ for the massless final state particles. 

In this paper we shall show that a similar rewriting is possible for the quantity $F^{\mu\nu}$ as
well. 
In particular, the expression for $F^{\mu\nu}$ can be manipulated into the form:
\ben \label{eaddcon4new}
  F^{\mu\nu} &  =&   {2\, G^3\over R\, c^{11}}\, \Bigg[ 4\,
\sum_{j=1}^{\tilde n} \wt p_j.n \sum_{\ell=1}^{\tilde n} \wt p_{\ell}.n 
\sum_{i=1}^{\tilde n} {\wt p_{i}^\mu \wt p_{i}^\nu\over \wt p_{i}.n} - 4\, 
\sum_{j=1}^{m} p'_j.n \sum_{\ell=1}^{m} p'_{\ell}.n \sum_{i=1}^{m} {p_{i}^{\prime\mu} p_{i}^{\prime\nu}\over 
p_{i}'.n}
\non\\ && \hskip -.5in  
+ 4\, \sum_{\ell=1}^{\tilde n} \wt p_{\ell}.n \sum_{i=1}^{\tilde n} \sum_{j=1\atop j\ne i}^{\tilde n}
 {1\over \wt p_{i}.n} {\wt p_{i}.\wt p_j\over \{(\wt p_{i}.\wt p_j)^2 - \wt p_{i}^2 \wt p_j^2\}^{3/2}} 
 \{2 (\wt p_{i}.\wt p_j)^2 - 3 \wt p_{i}^2 \wt p_j^2\} \{ n.\wt p_j \, \wt p_{i}^\mu \, \wt p_{i}^\nu - n.\wt p_{i} \, \wt p_{i}^\mu \, \wt p_j^\nu\}
 \non\\ &&   \hskip -.5in
+2\, \sum_{\ell=1}^{m} p'_{\ell}.n \sum_{i=1}^{m} \sum_{j=1\atop j\ne i}^{m}
 {1\over p'_{i}.n} {p'_{i}.p'_j\over \{(p'_{i}.p'_j)^2 - p_{i}^{\prime 2} p_j^{\prime 2}\}^{3/2}} 
 \{2 (p_{i}'.p_j')^2 - 3 p_{i}^{\prime 2} p_j^{\prime 2}\} \{ n.p'_j \, p_{i}^{\prime\mu} \, p_{i}^{\prime \nu} - 
 n.p'_{i} \, p_{i}^{\prime \mu} \, p_j^{\prime \nu}\} \non\\ &&   \hskip -.5in
+  \sum_{i=1}^{\tilde n} \sum_{j=1\atop j\ne i}^{\tilde n}
\sum_{\ell=1\atop \ell\ne i}^{\tilde n}  {1\over \wt p_{i}.n}  
{\wt p_{i}.\wt p_j\over \{(\wt p_{i}.\wt p_j)^2 - \wt p_{i}^2 \wt p_j^2\}^{3/2}}   
\{2 (\wt p_{i}.\wt p_j)^2 - 3 \wt p_{i}^2 \wt p_j^2\} {\wt p_{i}.\wt p_{\ell}\over \{(\wt p_{i}.\wt p_{\ell})^2 - \wt p_{i}^2 \wt p_{\ell}^2\}^{3/2}} \non\\ &&
\{2 (\wt p_{i}.\wt p_{\ell})^2 - 3 \wt p_{i}^2 \wt p_{\ell}^2\}  \{ n.\wt p_j \, \wt p_{i}^\mu  - n.\wt p_{i} \, \wt p_j^\mu \} \, 
\{ n.\wt p_{\ell} \, \wt p_{i}^\nu - n.\wt p_{i} \, \wt p_{\ell}^\nu \}\non \\
&& \hskip -.5in  + 4\, n.P' P^{\prime\mu} P^{\prime\nu} - 4\, n.\wt P \, \wt P^\mu \wt P^\nu \Bigg]\, .
\een
This corresponds to restricting the sum over final state particles in \refb{eaddcon4} 
to over massive particles
only, and adding the compensating term given in the last line of \refb{eaddcon4new}.
 
It follows from \refb{ebmnmod} and \refb{eaddcon4new} that if the final state contains only massless
particles, then $B_{\mu\nu}$ and $F_{\mu\nu}$ will depend on only the momenta of the objects
in the initial state. In fact it can also be seen with little effort that if the final state contains one massive
object and arbitrary number of massless objects, then $B_{\mu\nu}$ and $F_{\mu\nu}$
still depend only on the momenta of the objects
in the initial state. In particular the terms proportional to $\wt P^\mu\wt P^\nu$ in these
expressions exactly cancel the terms involving $\wt p_i$.

Finally we would like to mention that if some of the initial and /or 
final state particles are charged then we
also need to take into account the effect of long range 
electromagnetic interaction among these
particles. These effects have been studied in \cite{1808.03288,1912.06413,2008.04376}.
Following the same procedure that will be described in \S\ref{s2} and \S\ref{s3}, it
is easy to show that even in the presence of charged particles in the initial and the final states,
the late time gravitational wave-form at future null infinity continues to be independent of the
individual momenta of the final state massless particles. This has been demonstrated in
appendix \ref{sa}. 

In \S\ref{smassless} and \ref{sb} we apply our results to analyze gravitational
radiation
emitted during scattering
of massless particles and compare the results with those in 
\cite{1409.4555,1812.08137,1901.10986}.

\sectiono{Review of the analysis of the subleading term} \label{s2}

In this section we shall briefly review the analysis of the subleading term leading to \refb{ebmnmod}.
For this we divide the set of final momenta $\{p_i\}$ into the massive particle momenta
$\wt p_i$ and the massless particle momenta $\{\wh p_i\}$. We now use the fact that if either 
$p_i$ or $p_j$ represents a massless particle, then we have:
\be \label{e2.1}
{ p_{i}.p_{j}\over 
\{(p_{i}.p_{j})^2 
-p_{i}^2 p_{j}^2 \}^{3/2}} \, 
\left\{{3\over 2} p_{i}^2 p_{j}^2 - (p_{i}.p_{j})^2\right\} =1\, .
\ee
Therefore the first term in the expression for $B^{\mu\nu}$ given in \refb{e6.2a} 
may be written as:
\ben\label{efirsthere}
 &&\sum_{i=1}^n \sum_{j=1\atop j\ne i}^n 
{ p_{i}.p_{j}\over 
\{(p_{i}.p_{j})^2 
-p_{i}^2 p_{j}^2 \}^{3/2}} \, 
\left\{{3\over 2} p_{i}^2 p_{j}^2 - (p_{i}.p_{j})^2\right\}\, {p_{i}^\mu \over n.p_{i}}\,
 (n.p_{j}\, p_{i}^\nu - n.p_{i}\, p_{j}^\nu ) \non\\
 &=& \sum_{i=1}^{\tilde n} \sum_{j=1\atop j\ne i}^{\tilde n} 
{ \wt p_{i}.\wt p_{j}\over 
\{(\wt p_{i}.\wt p_{j})^2 
-\wt p_{i}^2 \wt p_{j}^2 \}^{3/2}} \, 
\left\{{3\over 2} \wt p_{i}^2 \wt p_{j}^2 - (\wt p_{i}.\wt p_{j})^2\right\}\, {\wt
p_{i}^\mu \over n.\wt p_{i}}\,
 (n.\wt p_{j}\, \wt p_{i}^\nu - n.\wt p_{i}\, \wt p_{j}^\nu ) \non\\
 && + \sum_{i} \sum_{j}
{\wt
p_{i}^\mu \over n.\wt p_{i}}\,
 (n.\wh p_{j}\, \wt p_{i}^\nu - n.\wt p_{i}\, \wh p_{j}^\nu ) 
 + \sum_{i} \sum_{j}
{\wh
p_{i}^\mu \over n.\wh p_{i}}\,
 (n.\wt p_{j}\, \wh p_{i}^\nu - n.\wh p_{i}\, \wt p_{j}^\nu ) \non\\
&& + \sum_{i} \sum_{j}
{\wh
p_{i}^\mu \over n.\wh p_{i}}\,
 (n.\wh p_{j}\, \wh p_{i}^\nu - n.\wh p_{i}\, \wh p_{j}^\nu ) \, .
\een
We can now use the result of momentum conservation:
\be
\sum_j \wt p_j^\nu + \sum_j \wh p_j^\nu = P^{\prime\nu}, \quad P'\equiv \sum_i p'_i\, ,
\ee
to express \refb{efirsthere} as,
\ben\label{esecondhere}
 &&\sum_{i=1}^n \sum_{j=1\atop j\ne i}^n 
{ p_{i}.p_{j}\over 
\{(p_{i}.p_{j})^2 
-p_{i}^2 p_{j}^2 \}^{3/2}} \, 
\left\{{3\over 2} p_{i}^2 p_{j}^2 - (p_{i}.p_{j})^2\right\}\, {p_{i}^\mu \over n.p_{i}}\,
 (n.p_{j}\, p_{i}^\nu - n.p_{i}\, p_{j}^\nu ) \non\\
  &=& \sum_{i=1}^{\tilde n} \sum_{j=1\atop j\ne i}^{\tilde n} 
{ \wt p_{i}.\wt p_{j}\over 
\{(\wt p_{i}.\wt p_{j})^2 
-\wt p_{i}^2 \wt p_{j}^2 \}^{3/2}} \, 
\left\{{3\over 2} \wt p_{i}^2 \wt p_{j}^2 - (\wt p_{i}.\wt p_{j})^2\right\}\, {\wt
p_{i}^\mu \over n.\wt p_{i}}\,
 (n.\wt p_{j}\, \wt p_{i}^\nu - n.\wt p_{i}\, \wt p_{j}^\nu ) \non\\
 && + \sum_{i} {\wt p_i^\mu \wt p_i^\nu\over n.\wt p_i}\, n.(P'-\wt P)
 + \sum_{i} {\wh p_i^\mu \wh p_i^\nu\over n.\wh p_i}\, n.P' + \wt P^\mu \wt P^\nu -
 P^{\prime \mu} P^{\prime \nu}\, .
\een
We also have
\ben \label{ethirdhere}
- \sum_{j=1}^n\, p_{j}.n  \sum_{i=1}^{n} \, {1\over p_{i}.n} \, p_{i}^\mu p_{i}^\nu
 &=& -n.P'\,   \sum_{i=1}^{n} \, {1\over \wt p_{i}.n} \, \wt p_{i}^\mu \wt p_{i}^\nu
  - n.P' \sum_i {1\over \wh p_{i}.n} \, \wh p_{i}^\mu \wh p_{i}^\nu\, .
 \een
 Substituting \refb{esecondhere} and \refb{ethirdhere} into \refb{e6.2a}, we get,
 \ben
 B^{\mu\nu} &=& -\, {4\, G^2\over R\, c^7} \left[
 \sum_{i=1}^{\tilde n} \sum_{j=1\atop j\ne i}^{\tilde n} 
{ \wt p_{i}.\wt p_{j}\over 
\{(\wt p_{i}.\wt p_{j})^2 
-\wt p_{i}^2 \wt p_{j}^2 \}^{3/2}} \, 
\left\{{3\over 2} \wt p_{i}^2 \wt p_{j}^2 - (\wt p_{i}.\wt p_{j})^2\right\}\, {\wt
p_{i}^\mu \over n.\wt p_{i}}\,
 (n.\wt p_{j}\, \wt p_{i}^\nu - n.\wt p_{i}\, \wt p_{j}^\nu )  \right.
\non \\ &&  \left.  - \left\{
 \sum_{j=1}^{\tilde n}\, \wt p_{j}.n  \sum_{i=1}^{\tilde n} \, {1\over \wt p_{i}.n} \, \wt p_{i}^\mu \wt 
 p_{i}^\nu
 -  \sum_{j=1}^m\, p_{j}'.n \sum_{i=1}^{m} \, {1\over p_{i}'.n} \, p_{i}^{\prime\mu} p_{i}^{\prime \nu}\right\}
 + \wt P^\mu \wt P^\nu -
 P^{\prime \mu} P^{\prime \nu}
 \right]\, .  \label{e6.2balt}
\een
This reproduces \refb{ebmnmod}.

\sectiono{Analysis of the subsubleading term} \label{s3}

We shall now rewrite  the expression \refb{eaddcon4} for $F^{\mu\nu}$ by dividing the sum over final
state momenta into the contribution from massless and massive particle momenta, denoted by
$\wh p_i$ and $\wt p_i$ respectively. $F_{(n)}^{\mu\nu}$ will denote contribution from terms
where in each term, we have $n$ factors of $\wh p_i$. Therefore we have:
\ben \label{e3.1}
  F_{(0)}^{\mu\nu} &  =&   {2\, G^3\over R\, c^{11}}\, \Bigg[ 4\,
\sum_{j=1}^{\tilde n} \wt p_j.n \sum_{\ell=1}^{\tilde n} \wt p_{\ell}.n 
\sum_{i=1}^{\tilde n} {\wt p_{i}^\mu \wt p_{i}^\nu\over \wt p_{i}.n} - 4\, 
\sum_{j=1}^{m} p'_j.n \sum_{\ell=1}^{m} p'_{\ell}.n \sum_{i=1}^{m} {p_{i}^{\prime\mu} p_{i}^{\prime\nu}\over 
p_{i}'.n}
\non\\ && \hskip -.5in  
+ 4\, \sum_{\ell=1}^{\tilde n} \wt p_{\ell}.n \sum_{i=1}^{\tilde n} \sum_{j=1\atop j\ne i}^{\tilde n}
 {1\over \wt p_{i}.n} {\wt p_{i}.\wt p_j\over \{(\wt p_{i}.\wt p_j)^2 - \wt p_{i}^2 \wt p_j^2\}^{3/2}} 
 \{2 (\wt p_{i}.\wt p_j)^2 - 3 \wt p_{i}^2 \wt p_j^2\} \{ n.\wt p_j \, \wt p_{i}^\mu \, \wt p_{i}^\nu - n.\wt p_{i} \, \wt p_{i}^{\mu} \, \wt p_j^{\nu}\}
 \non\\ &&   \hskip -.5in
+2\, \sum_{\ell=1}^{m} p'_{\ell}.n \sum_{i=1}^{m} \sum_{j=1\atop j\ne i}^{m}
 {1\over p'_{i}.n} {p'_{i}.p'_j\over \{(p'_{i}.p'_j)^2 - p_{i}^{\prime 2} p_j^{\prime 2}\}^{3/2}} 
 \{2 (p_{i}'.p_j')^2 - 3 p_{i}^{\prime 2} p_j^{\prime 2}\} \{ n.p'_j \, p_{i}^{\prime\mu} \, p_{i}^{\prime \nu} - 
 n.p'_{i} \, p_{i}^{\prime \mu} \, p_j^{\prime \nu}\} \non\\ &&   \hskip -.5in
+  \sum_{i=1}^{\tilde n} \sum_{j=1\atop j\ne i}^{\tilde n}
\sum_{\ell=1\atop \ell\ne i}^{\tilde n}  {1\over \wt p_{i}.n}  
{\wt p_{i}.\wt p_j\over \{(\wt p_{i}.\wt p_j)^2 - \wt p_{i}^2 \wt p_j^2\}^{3/2}}   
\{2 (\wt p_{i}.\wt p_j)^2 - 3 \wt p_{i}^2 \wt p_j^2\} {\wt p_{i}.\wt p_{\ell}\over \{(\wt p_{i}.\wt p_{\ell})^2 - \wt p_{i}^2 \wt p_{\ell}^2\}^{3/2}} \non\\ &&
\{2 (\wt p_{i}.\wt p_{\ell})^2 - 3 \wt p_{i}^2 \wt p_{\ell}^2\}  \{ n.\wt p_j \, \wt p_{i}^\mu  - n.\wt p_{i} \, \wt p_j^\mu \} \, 
\{ n.\wt p_{\ell} \, \wt p_{i}^\nu - n.\wt p_{i} \, \wt p_{\ell}^\nu \}\Bigg]\, .
\een
In writing down the expressions for $F^{\mu\nu}_{(n)}$ for $n\ge 1$, we shall try to simplify the
sum over final state momenta using the relations:
\be
\sum_j \wt p_j^\mu=\wt P^\mu, \qquad \sum_j \wh p_j^\mu = (P'-\wt P)^\mu\, .
\ee
Another simplification follows from the observation that the expression
$\{ n. p_j \,  p_{i}^\mu \,  p_{i}^\nu - n. p_{i} \,  p_{i}^{\mu} \,  p_j^{\nu}\}$
vanishes for $j=i$. Therefore unless the factor multiplying it diverges for $j=i$, we can include
in the sum over $j$ the term $j=i$ even if the original sum excludes this. This trick can often 
be used to 
make the sum over $i$ and $j$ into independent sums as long as either $i$ or $j$ represents a
massless particle, since in this case we can first use \refb{e2.1} to replace the apparently divergent 
factor at $j=i$ by a finite term, and then include the contribution from the $j=i$ term in the sum.
This gives:
\ben\label{e3.2}
F_{(1)}^{\mu\nu} &  =&   {2\, G^3\over R\, c^{11}}\, \Bigg[ 8\, n.(P'-\wt P) \,
 n.\wt P\,  
\sum_{i=1}^{\tilde n} {\wt p_{i}^\mu \wt p_{i}^\nu\over \wt p_{i}.n}
+ 4\, n.\wt P \, n.\wt P \, \sum_i {\wh p_{i}^\mu \wh p_{i}^\nu\over \wh p_{i}.n} \non\\ 
&&\hskip -.5in 
+\, 4\, n.(P'-\wt P) \, \sum_{i} \sum_{j\ne i}
 {1\over \wt p_{i}.n} {\wt p_{i}.\wt p_j\over \{(\wt p_{i}.\wt p_j)^2 - \wt p_{i}^2 \wt p_j^2\}^{3/2}} 
 \, \{2 (\wt p_{i}.\wt p_j)^2 - 3 \wt p_{i}^2 \wt p_j^2\} \{ n.\wt p_j \, \wt p_{i}^\mu \, \wt p_{i}^\nu - n.\wt p_{i} \, \wt p_{i}^{\mu} \, \wt p_j^{\nu}\} \non \\
 &&  -\, 8\, n.\wt P\,  \sum_{i} {1\over \wt p_i.n} 
 \{ n.(P'-\wt P) \, \wt p_{i}^\mu \, \wt p_{i}^\nu - n.\wt p_{i} \, \wt p_{i}^{(\mu} \, (P'-\wt P)^{\nu)}\} \non\\ &&
 - 8\, n.\wt P\,  \sum_{i} {1\over \wh p_i.n} 
 \{ n.\wt P \, \wh p_{i}^\mu \, \wh p_{i}^\nu - n.\wh p_{i} \, \wh p_{i}^{(\mu} \, \wt P^{\nu)}\} \non\\ &&
 +\, 4\, \sum_{i} {1\over \wh p_i.n}\, \{ n.\wt P \, \wh p_{i}^\mu  - n.\wh p_{i} \, \wt P^\mu \} \, 
\{ n.\wt P \, \wh p_{i}^\nu - n.\wh p_{i} \, \wt P^\nu \} \non\\ &&
-\, 4 \sum_i \sum_{j\ne i}  {1\over \wt p_{i}.n}  %
{\wt p_{i}.\wt p_j\over \{(\wt p_{i}.\wt p_j)^2 - \wt p_{i}^2 \wt p_j^2\}^{3/2}}   
\{2 (\wt p_{i}.\wt p_j)^2 - 3 \wt p_{i}^2 \wt p_j^2\} \non\\ && 
\bigg\{ n.\wt p_j \, n.(P'-\wt P) \, \wt p_{i}^\mu \wt p_{i}^\nu - n.\wt p_{i} \, n.(P'-\wt P)\, \wt p_j^{\mu} 
\, \wt p_{i}^{\nu}
- n.\wt p_{j} \, n.\wt p_{i}\, \wt p_i^{(\mu} 
\, (P'-\wt P)^{\nu)}  \non\\ &&
\hskip 1in + (n.\wt p_{i})^2 \wt p_j^{(\mu} 
\, (P'-\wt P)^{\nu)}
 \bigg\}\Bigg]\, .
\een
This can be simplified to:
\ben\label{e3.2apre}
F_{(1)}^{\mu\nu} &  =&   {2\, G^3\over R\, c^{11}}\, \Bigg[ 
4\, n.\wt P\, (P'-\wt P)^{\mu} \, \wt P^{\nu} + 4\, n.\wt P\, (P'-\wt P)^{\nu} \, \wt P^{\mu}
+ \, 4 \, n.(P'-\wt P)\, \wt P^\mu \wt P^\nu
\non\\ &&
-\, 4 \sum_i \sum_{j\ne i} 
{\wt p_{i}.\wt p_j\over \{(\wt p_{i}.\wt p_j)^2 - \wt p_{i}^2 \wt p_j^2\}^{3/2}}   
\{2 (\wt p_{i}.\wt p_j)^2 - 3 \wt p_{i}^2 \wt p_j^2\} \non\\ && 
\hskip .3in \times\, \bigg\{ - n.\wt p_{j} \, \wt p_i^{(\mu} 
\, (P'-\wt P)^{\nu)}  + n.\wt p_{i} \wt p_j^{(\mu} 
\, (P'-\wt P)^{\nu)}
 \bigg\}\Bigg]\, .
\een
We now note that the last line is anti-symmetric under the exchange of $i$ and $j$ while the second
line is symmetric under this exchange. Therefore this term vanishes after summing over $i$ and $j$,
and we get:
\be\label{e3.2a}
F_{(1)}^{\mu\nu}   =   {2\, G^3\over R\, c^{11}}\, \Bigg[ 
4\, n.\wt P\, (P'-\wt P)^{\mu} \, \wt P^{\nu} + 4\, n.\wt P\, (P'-\wt P)^{\nu} \, \wt P^{\mu}
+ \, 4 \, n.(P'-\wt P)\, \wt P^\mu \wt P^\nu
\Bigg]\, .
\ee
We also have,
\ben\label{e3.3}
F_{(2)}^{\mu\nu} &  =&   {2\, G^3\over R\, c^{11}}\, \Bigg[4\,
n.(P'-\wt P) \, n.(P'-\wt P)
\sum_{i=1}^{\tilde n} {\wt p_{i}^\mu \wt p_{i}^\nu\over \wt p_{i}.n}
+ 8\, n.(P'-\wt P) \, n.\wt P\, 
\sum_{i} {\wh p_{i}^\mu \wh p_{i}^\nu\over \wh p_{i}.n} \non\\ &&
 - 8\, n.(P'-\wt P)\, \sum_i  {1\over \wh p_{i}.n} \{ n.\wt P \, \wh p_{i}^\mu \, \wh p_{i}^\nu - n.\wh p_{i} \, 
 \wh p_{i}^{\mu} \, \wt P^{\nu}\}\non\\ &&
  - 8\, n.(P'-\wt P)\, \sum_i  {1\over \wt p_{i}.n} \bigg\{ n.(P'-\wt P) \, \wt p_{i}^\mu \, \wt p_{i}^\nu 
  - n.\wt p_{i} \, 
 \wt p_{i}^{\mu} \, (P'-\wt P)^{\nu}\bigg\}\non\\ &&
  - 8\, n.\wt P\, \sum_i  {1\over \wh p_{i}.n} \bigg\{ n.(P'-\wt P) \, \wh p_{i}^\mu \, \wh p_{i}^\nu 
   - n.\wh p_{i} \, 
 \wh p_{i}^{\mu} \, (P'-\wt P)^{\nu} \bigg\} \non\\ &&
 + 4\, \sum_i {1\over \wt p_i.n}\, \bigg\{ n.(P'-\wt P) \, n.(P'-\wt P) \, \wt p_{i}^\mu \wt p_{i}^\nu 
 - n.\wt p_{i} \, n.(P'-\wt P)\, (P'-\wt P)^{\mu} 
\, \wt p_{i}^{\nu} \non\\ &&
- n.(P'-\wt P) \, n.\wt p_{i}\, \wt p_i^{\mu} 
\, (P'-\wt P)^{\nu} 
 + (n.\wt p_{i})^2 (P'-\wt P)^{\mu} 
\, (P'-\wt P)^{\nu}
 \bigg\} \non\\ &&
 +\, 8\, \sum_{i} {1\over \wh p_i.n}\, \{ n.(P'-\wt P) \, \wh p_{i}^{(\mu} 
  - n.\wh p_{i} \, (P'-\wt P)^{(\mu} \} \, 
\{ n.\wt P \, \wh p_{i}^{\nu)} - n.\wh p_{i} \, \wt P^{\nu)} \}
\Bigg]\, .
\een
This can be simplified to:
\ben\label{e3.3a}
F_{(2)}^{\mu\nu} &  =&   {2\, G^3\over R\, c^{11}}\, \Bigg[ 8\, n.(P'-\wt P)\,
(P'-\wt P)^{(\mu} \, \wt P^{\nu)}
  + 4\,  n.\wt P\, (P'-\wt P)^{\mu} 
\, (P'-\wt P)^{\nu}
\Bigg] \, .
\een
Finally we have,
\ben\label{e3.4}
F_{(3)}^{\mu\nu} &  =&   {2\, G^3\over R\, c^{11}}\, \Bigg[4\,
n.(P'-\wt P) \, n.(P'-\wt P)
\sum_i {\wh p_{i}^\mu \wh p_{i}^\nu\over \wh p_{i}.n}\non\\ &&
 - 8\, n.(P'-\wt P)\, \sum_i  {1\over \wh p_{i}.n} \{ n.(P'-\wt P) \, \wh p_{i}^\mu \, \wh p_{i}^\nu 
 - n.\wh p_{i} \, 
 \wh p_{i}^\mu \, (P'-\wt P)^\nu\}  \non\\ &&
 +\, \,4 \sum_{i} {1\over \wh p_i.n}\, \{ n.(P'-\wt P) \, \wh p_{i}^{\mu} 
  - n.\wh p_{i} \, (P'-\wt P)^{\mu} \} \non\\ && \hskip 1in
\{ n.(P'-\wt P) \, \wh p_{i}^{\nu} - n.\wh p_{i} \, (P'-\wt P)^{\nu} \}
\Bigg]\, .
\een
This can be simplified to:
\ben\label{e3.4a}
F_{(3)}^{\mu\nu} &  =&   {2\, G^3\over R\, c^{11}}\, \Bigg[
  4 \, n.(P'-\wt P) (P'-\wt P)^\mu (P'-\wt P)^{\nu}
\Bigg]\, .
\een
Using \refb{e3.2a}, \refb{e3.3a} and \refb{e3.4a}, we get,
\be \label{e3.6}
F_{(1)}^{\mu\nu}+F_{(2)}^{\mu\nu}+F_{(3)}^{\mu\nu} = {8\, G^3\over R\, c^{11}}\, \left[ 
n.P'\, P^{\prime\mu} \, P^{\prime\nu} - n.\wt P\, \wt P^\mu \wt P^\nu\right]\, .
\ee
Adding this to \refb{e3.1} we get \refb{eaddcon4new}.
\def\vx{\vec x}

\sectiono{Example involving scattering of massless particles} \label{smassless}

In this section we shall compare our results to that of \cite{1409.4555,1812.08137} 
on the emission
of soft gravitational
radiation during
the scattering of a pair of massless particles. For this comparison we shall set $c=1$
since the results of \cite{1812.08137} were given in that convention.\footnote{A similar
result was found in \cite{1901.10986} where part of the contribution associated to the
Coulomb phase of the soft radiation 
({\it e.g.} the terms in the second line of \refb{e6.2a}) was not included. 
}
Let $\wt e_{\mu\nu}(\omega,\vx)$ be the time Fourier transform of $e_{\mu\nu}(t,\vx)$:
\be 
\wt e_{\mu\nu}(\omega,\vx) = \int du \, e^{i\omega u} \, e_{\mu\nu}(t, \vx), \qquad u=t-t_0\, .
\ee
Then $\wt e_{\mu\nu}(\omega,\vx)$ has a small $\omega$ expansion of the 
form\cite{1912.06413}:
\be \label{e4.0}
\wt e_{\mu\nu}(\omega,\vx) = i\, A_{\mu\nu}\, \omega^{-1} - (B_{\mu\nu}-C_{\mu\nu})\ln\omega
+{i\over 2} (F_{\mu\nu}-G_{\mu\nu})\, \omega (\ln\omega)^2 + \cdots\, .,,
\ee
where $A_{\mu\nu}$, $B_{\mu\nu}$, $C_{\mu\nu}$, $F_{\mu\nu}$ and $G_{\mu\nu}$ are the
same coefficients that appeared in the large $|u|$ expansion of $e_{\mu\nu}$.
The $i\eps$ prescription inside $\ln\omega$ captures separate information on $B_{\mu\nu}$ and
$C_{\mu\nu}$ and also on $F_{\mu\nu}$ and $G_{\mu\nu}$\cite{1912.06413}, but 
at present we shall proceed ignoring the $i\eps$ prescription.

When all the incoming and outgoing particles are massless, we get from \refb{e6.2b},
\refb{eaddcon5}, \refb{ebmnmod} and \refb{eaddcon4new},
\be \label{e4.1}
B^{\mu\nu}=-C^{\mu\nu} = -{4\, G^2\over R} \left[ n.P'\, \sum_i {p_i^{\prime\mu}p_i^{\prime\nu}
\over n.p'_i} - P^{\prime\mu} P^{\prime\nu}\right]\, ,
\ee
and
\be \label{e4.2}
F^{\mu\nu}=-G^{\mu\nu} =-{16 \, G^3\over R} \, n.P' \, 
\left[ n.P'\, \sum_i {p_i^{\prime\mu}p_i^{\prime\nu}
\over n.p'_i} - P^{\prime\mu} P^{\prime\nu}\right]\, .
\ee
We also have, from \refb{e6.2},
\be \label{e6.2repeat}
A^{\mu\nu} = {2\, G\over R} \, \left[-\sum_{i=1}^{n}  
p_{i}^\mu \, p_{i}^\nu\, 
{1\over n.p_{i}} + \sum_{i=1}^{m}  
p_{i}^{\prime\mu} \, p_{i}^{\prime\nu}\, 
{1\over n.p_{i}'}  \right]\, .
\ee

We now apply these results to the specific case of scattering of two massless particles 
into two massless
particles and  soft gravitational radiation. Following  \cite{1812.08137}, we label 
the momenta of the incoming and outgoing 
hard particles as:
\ben
&& p_1'=E(1,0,0,1), \quad p_2'=E(1,0,0,-1), \non\\ && 
p_1=E(1,\sin\Theta_s\cos\phi, \sin\Theta_s\sin\phi,\cos\Theta_s), \non\\ &&
p_2=E(1,-\sin\Theta_s\cos\phi,-\sin\Theta_s\sin\phi, -\cos\Theta_s)\, .
\een
On the other hand, the direction of emission of the soft gravitational rediation, encoded in the
four vector $n=(1,\hat n)$, takes the form:
\be
n = (1,\sin\theta, 0, \cos\theta)\, .
\ee
Our choice of frame is rotated by an angle $\phi$ about the $z$-axis relative to the frame used
in \cite{1812.08137}, so that the direction of propagation of the soft gravitational wave, and not 
the momenta of the outgoing
hard particles, lies in the $x$-$z$ plane. We now define
\be
\hat e^+ = (0,\cos\theta, i, -\sin\theta), \quad \hat e^- = (0,\cos\theta, -i, -\sin\theta)\, ,
\ee
so that,
\be
\eps^+_{\mu\nu} ={1\over 2} \, \hat e^+_{\mu}\hat e^+_{\nu}, \qquad
\eps^-_{\mu\nu} ={1\over 2} \, \hat e^-_{\mu}\hat e^-_{\nu}\, ,
\ee
denote left and right circular polarizations of soft gravitational waves traveling along $n$.

Using \refb{e4.0} we now get,
\be \label{e4.10a}
\eps^\pm_{\mu\nu}  \, \wt e^{\mu\nu}(\omega,\vx)= \eps^\pm_{\mu\nu} 
\left[ i\, A^{\mu\nu}\, \omega^{-1} 
- (B^{\mu\nu}-C^{\mu\nu})\ln\omega
+{i\over 2} (F^{\mu\nu}-G^{\mu\nu})\, \omega (\ln\omega)^2 + \cdots
\right]\, .
\ee
Since \cite{1812.08137} gives the result for small $\theta$ and $\Theta_s$, we shall also
make this approximation.
However, in \S\ref{sb} we have given the results
for finite $\theta$ and $\Theta_s$.
Now, for small $\theta,\Theta_s$,
\ben \label{eresults}
&&
 \hat e^\pm . p_1'=-E\, \sin\theta\simeq -E\,\theta, \qquad \hat e^\pm.p_2'=E\,\sin\theta
 \simeq E\, \theta, \non\\ && 
 \hat e^\pm . p_1=
E\, (\sin\Theta_s \cos\theta\cos\phi-\cos\Theta_s\sin\theta\pm i\, \sin\Theta_s\sin\phi)
\simeq E(\Theta_s\, \cos\phi-\theta \pm i\Theta_s\sin\phi)\, ,
\non\\ &&
\hat e^\pm . p_2=-
E\, (\sin\Theta_s \cos\theta\cos\phi-\cos\Theta_s\sin\theta\pm i\, \sin\Theta_s\sin\phi)
\simeq -E(\Theta_s\, \cos\phi-\theta \pm i\Theta_s\sin\phi)\, , \non\\ &&
n.p_1'=-E\, (1-\cos\theta)\simeq - E\, \theta^2/2, \quad n.p_2'=-E\, (1+\cos\theta)
\simeq -2\, E\non\\ &&
n.p_1 = -E\, (1-\sin\Theta_s \sin\theta\cos\phi-\cos\Theta_s\cos\theta)
\simeq - E\, (\Theta_s^2 + \theta^2 - 2\, \Theta_s \theta\, \cos\phi)/2\non\\ &&
n.p_2 = -E\, (1+\sin\Theta_s \sin\theta\cos\phi+\cos\Theta_s\cos\theta)\simeq - 2\, E\, .
\een
It follows from these equations that $\eps^\pm_{\mu\nu} p_i^\mu p_i^\nu$ and
$\eps^\pm_{\mu\nu} p_i^{\prime\mu} p_i^{\prime\nu}$ are quadratic in the small parameters 
$\Theta_s$ and $\theta$. Therefore only terms with $n.p_1$ or $n.p_1'$ in the denominator
will survive in this limit. This gives:
\ben\label{eAmunu}
\eps^\pm_{\mu\nu} A^{\mu\nu} &=&
{G\over R} \, \left[ -
{\hat e^\pm.p_1 \, \hat e^\pm.p_1
\over n.p_1} + {\hat e^\pm.p_1' \, \hat e^\pm.p_1'
\over n.p'_1} \right]\non\\ &=&
{2\, G\, E\over R} \, \left[
{(\Theta_s\, \cos\phi-\theta \pm i\Theta_s\sin\phi)^2\over 
 (\Theta_s^2 + \theta^2 - 2\, \Theta_s \theta\, \cos\phi)}
 -1
 \right]\, ,
\een 
\be\label{efin1}
\eps^\pm_{\mu\nu} (B^{\mu\nu}-C^{\mu\nu}) =-{4\, G^2\over R} \,  n.P'\, 
{\hat e^\pm.p_1' \, \hat e^\pm.p_1'
\over n.p'_1} =-{16\, G^2 E^2\over R}\, ,
\ee
and
\be\label{efin2}
\eps^\pm_{\mu\nu} (F^{\mu\nu}-G^{\mu\nu}) =-{16\, G^3\over R} \,  (n.P')^2\, 
{\hat e^\pm.p_1' \, \hat e^\pm.p_1'
\over n.p'_1} ={128\, G^3 E^3\over R}\, .
\ee
Following \cite{1812.08137},  we introduce the 
variable $\psi$ via:
\be
\sin\psi = {\Theta_s\sin\phi\over  (\Theta_s^2 + \theta^2 - 2\, \Theta_s \theta\, \cos\phi)^{1/2}},
\qquad \cos\psi = {\Theta_s\cos\phi-\theta \over  (\Theta_s^2 + \theta^2 - 2\, \Theta_s \theta\, \cos\phi)^{1/2}
}\, ,
\ee
so that \refb{eAmunu} may be expressed as:
\be\label{eAmununew}
\eps^\pm_{\mu\nu} A^{\mu\nu} = {2\, G\, E\over R} \, \left[ e^{\pm 2\, i\, \psi}-1\right]\, .
\ee
Substituting \refb{efin1}, \refb{efin2} and \refb{eAmununew} into \refb{e4.10a}, we get,
\be \label{efin}
\eps^\pm_{\mu\nu} \wt e^{\mu\nu} = i\, {2\, G\, E\over R}\, \omega^{-1}\, \left[
e^{\pm 2i\psi}-1 - i\, 8\, G\, E\, \omega\, \ln\omega + 32\, G^2 \, E^2 \, \omega^2\, (\ln\omega)^2
\right]\, .
\ee
Up to an overall normalization this agrees with the small $\omega$ expansion of eq.(6.20) of
\cite{1812.08137} after identifying the variable $R$ of \cite{1812.08137}, describing the
Schwarzschild radius of the system, with $4GE$. We shall now verify that the overall
normalization also agrees.

To check the overall normalization, we compute the energy flux associated with
\refb{efin} at the leading order in $\omega$. 
This can be done using the formula for the angular distribution of the
energy flux with a given 
polarization. In the $8\pi G=1$ unit the flux is given by (see {\it e.g.} \cite{1801.07719}):
\be \label{ecomp1}
{d E_\pm\over d\omega d\Omega} = {\omega^2\over\pi} \, R^2 \, |\eps^\pm_{\mu\nu} \wt e^{\mu\nu} |^2
= {E^2\over 16\pi^3}\left| e^{\pm 2i\psi}-1\right|^2\, .
\ee
On the other hand, the same flux computed in \cite{1812.08137} at the leading order in
$\omega$ is given by (see eq.(6.13)):
\be\label{ecomp2}
{4\, G\, E^2\over 8\pi^2} \left| e^{\pm 2i\psi}-1\right|^2={E^2\over 16\pi^3}
\left| e^{\pm 2i\psi}-1\right|^2\, .
\ee
Comparing \refb{ecomp1} and \refb{ecomp2} we see that the overall normalizations also
match.

Even though we have derived the various formul\ae\ in the limit of small $\Theta_s$ and $\theta$
for comparison with the results of \cite{1812.08137}, it follows from our general result that even for
general values of $\Theta_s$ and $\theta$, our expressions for $B_{\mu\nu}=-C_{\mu\nu}$ and
$F_{\mu\nu}=-G_{\mu\nu}$ remain the same as those given in \refb{efin1} and \refb{efin2}. 
What is perhaps more striking is that even if the incoming states have a small enough impact 
parameter so that they form a black hole, possibly accompanied by hard radiation, the expressions
for $B_{\mu\nu}$ and $F_{\mu\nu}$ do not change. This follows from the discussion in the
last paragraph of \S\ref{s1} since we have only one massive object in the final state.

Before concluding this section, we would like to 
discuss another aspect of the results given in \cite{1812.08137}. 
\cite{1812.08137} used the wave-form \refb{efin} to compute the total flux of
soft radiation to subsubleading order. However since in principle there could be terms
of order $\omega$ inside the square bracket that have not been computed, it could
give an additional contribution at the subleading order, spoiling the subsubleading results
of \cite{1812.08137}. It is easy to see that an imaginary term of order $\omega$ inside the
square bracket in \refb{efin} will not contribute to the energy flux at the subleading order after
summing over polarizations of the gravitational radiation. On the other hand a real  term proportional to
$\omega$ will violate the reality condition $\wt e_{\mu\nu}(\omega)^*=\wt e_{\mu\nu}(-\omega)$ that
is required for the reality of the gravitational field. However there could be a contribution proportional
to $\omega\{H(\omega)-H(-\omega)\}$ inside the square bracket, with $H$ denoting the Heaviside
function, that satisfies the reality 
condition. If present, such a term would give subleading contribution to the energy flux,
spoiling the result of \cite{1812.08137}.

We shall now show that such a term is absent, but for this we need to carefully keep track
of the 
$i\eps$ prescription in
the argument of the logarithms in \refb{e4.0}. 
It follows from the analysis of \cite{1912.06413}, that 
at the subleading
order, the  time Fourier transform of the wave-form, including the $i\eps$ prescription, is 
given by:
\be
- {1\over 2} \left\{ B_{\mu\nu}\ln(\omega+i\eps) -C_{\mu\nu}\ln(\omega-i\eps)
+ B_{\mu\nu}\ln(-\omega-i\eps) -C_{\mu\nu}\ln(-\omega+i\eps))
\right\}\, .
\ee
Using \refb{e4.1} this can be written as:
\be
{1\over 2} C_{\mu\nu} \{ \ln(\omega+i\eps)+\ln(\omega-i\eps)+\ln(-\omega-i\eps)
+ \ln(-\omega+i\eps)\} = 2\, C_{\mu\nu} \, \ln|\omega|\, .
\ee
Therefore at the subleading order we do not have terms proportional to
$H(\omega)-H(-\omega)$. 
Note that this is a consequence of the
relation $B_{\mu\nu}=-C_{\mu\nu}$ and seems to be present when all the incoming and
outgoing particles are massless. From this it follows that \refb{efin} can be used to compute the
total flux of energy carried by the gravitational radiation
to order $\omega^2(\ln\omega)^2$, reproducing the result of
\cite{1812.08137}. 

\sectiono{Energy flux from massless particle scattering} \label{sb}

In \S\ref{smassless} we compared our results for radiation during scattering of massless particles
at small angle with those of \cite{1812.08137}.
In this section we shall compute the energy flux of low frequency gravitational radiation 
produced during such a scattering without
making the small angle approximation.

Using \refb{e4.1}-\refb{e6.2repeat} and \refb{eresults} without making the small $\theta,\Theta_s$
approximation, we get

\ben\label{eAmununewer}
&& \eps^\pm_{\mu\nu} A^{\mu\nu} = {2\, G\, E\over R} \, \left[ e^{\pm 2\, i\, \psi}-1\right]\, ,
\nonumber \\ &&
\sin\psi={\sin\Theta_s\sin\phi\over 
\{1 - (\sin\Theta_s \sin\theta\cos\phi+\cos\Theta_s\cos\theta)^2\}^{1/2}},
\nonumber \\ && \cos\psi = {\sin\Theta_s \cos\theta\cos\phi-\cos\Theta_s\sin\theta \over 
\{1 - (\sin\Theta_s \sin\theta\cos\phi+\cos\Theta_s\cos\theta)^2\}^{1/2}},
\een
\be \label{eb2}
\eps^\pm_{\mu\nu} (B^{\mu\nu}-C^{\mu\nu})=-{16\, G^2 E^2\over R}\, ,
\qquad
\eps^\pm_{\mu\nu} (F^{\mu\nu}-G^{\mu\nu})={128\, G^3 E^3\over R}\, .
\ee
Using \refb{e4.0} and the first equality in \refb{ecomp1}, and summing over polarizations, 
we now get the total differential flux in the $8\pi G=1$ unit:
\be
{d E\over d\omega d\Omega}\equiv \sum_\pm {d E_\pm\over d\omega d\Omega} = \sum_\pm 
{\omega^2\over\pi} \, R^2 \, \left|\eps^\pm_{\mu\nu} \left\{i\, A^{\mu\nu}\, \omega^{-1} 
- \ln\omega (B^{\mu\nu}-C^{\mu\nu}) +{i\over 2} \omega (\ln\omega)^2 (F^{\mu\nu}-G^{\mu\nu}) 
\right\}
\right|^2
\, .
\ee
Using \refb{eAmununewer}, \refb{eb2}, this can be written as
\be\label{efluxdiff}
P(\theta,\phi) + Q(\theta,\phi)\, \omega^2(\ln\omega)^2 + \OO(\omega^2\ln\omega)\, ,
\ee
where
\be\label{edefpq}
P(\theta,\phi) = {E^2\over 2\pi^3} \sin^2\psi, \qquad 
Q(\theta,\phi)= {E^4\over 8\, \pi^5}\, (1-2\sin^2\psi)\, .
\ee
Therefore for computing the total flux, we need to evaluate the integral:
\be
I\equiv 
\int_0^\pi \sin\theta d\theta \int_0^{2\pi} d\phi\, \sin^2 \psi 
=\int_0^\pi \sin\theta d\theta \int_0^{2\pi} d\phi\, {\sin^2\Theta_s\sin^2\phi\over 
\{1 - (\sin\Theta_s \sin\theta\cos\phi+\cos\Theta_s\cos\theta)^2\}}\, .
\ee
Now, writing
\ben
{\sin^2\Theta_s\sin^2\phi\over 
\{1 - (\sin\Theta_s \sin\theta\cos\phi+\cos\Theta_s\cos\theta)^2\}}
&=& {1\over 2} {\sin^2\Theta_s\sin^2\phi\over 
\{1 - (\sin\Theta_s \sin\theta\cos\phi+\cos\Theta_s\cos\theta)\}} \nonumber \\
&& + (\theta\to\pi-\theta, \phi\to\phi+\pi)\, ,
\een
and noting that both terms produce the same integral, we can express $I$ as:
\be\label{eintegral}
I = \int_0^\pi \sin\theta d\theta \int_0^{2\pi} d\phi\, {\sin^2\Theta_s\sin^2\phi\over 
\{1 - (\sin\Theta_s \sin\theta\cos\phi+\cos\Theta_s\cos\theta)\}}\, .
\ee
We can carry out the $\phi$ integral by defining $z=e^{i\phi}$, 
replacing $\sin\phi$ by $(z-z^{-1})/(2i)$, $\cos\phi$ by $(z+z^{-1})/2$, and
regarding \refb{eintegral} as
a contour integral over $z$ along the unit circle. 
The resulting integrand has double pole at $z=0$ and single poles at $z=z_\pm$
where 
\be
z_{\pm}=\frac{(1\pm \cos\theta_s)(1\mp \cos\theta)}{\sin\theta\sin\theta_s}\, .
\ee
It is easy to see that for $\theta>\Theta_s$, $z_+>1$, $z_-<1$ and for 
$\theta<\Theta_s$, $z_+<1$,  $z_->1$. 
We can now perform the contour integral by picking up the residues at the poles 
inside the unit circle. 
The result is:
\be
I = 2\pi\, \int_0^\pi \sin\theta d\theta \left[ H(\Theta_s-\theta) {1+\cos\Theta_s\over
1+\cos\theta} + H(\theta-\Theta_s) {1-\cos\Theta_s\over 1-\cos\theta}\right]\, ,
\ee
where $H$ is the step function. This arises due to the fact that as $\theta$ varies from being
below $\Theta_s$ to above $\Theta_s$, the poles of the integrand move across the
integration contour. After carrying out the $\theta$ integration we get:
\be
I = 2\pi\, \left[2\, \ln\, 2 - (1+\cos\Theta_s)\ln(1+\cos\Theta_s) - (1-\cos\Theta_s)\ln(1-\cos\Theta_s)
\right]\, .
\ee
We can now use \refb{edefpq} to calculate the energy flux integrated over all angles, up to order 
$\omega^2(\ln\omega)^2$:
\ben \label{eb16}
&& \int  {d E\over d\omega d\Omega} d\Omega
=\int_0^\pi \sin\theta d\theta \int_0^{2\pi} d\phi\, [P(\theta,\phi) + Q(\theta,\phi)\, \omega^2(\ln\omega)^2]
\nonumber \\ 
&=& {E^2\over \pi^2} \left[2\, \ln\, 2 - (1+\cos\Theta_s)\ln(1+\cos\Theta_s) - (1-\cos\Theta_s)\ln(1-\cos\Theta_s)
\right] \nonumber \\ &+&
{E^4\over 2\pi^4} \omega^2 (\ln\omega)^2 \left[1 - 2\, \ln\, 2 
+ (1+\cos\Theta_s)\ln(1+\cos\Theta_s) + (1-\cos\Theta_s)\ln(1-\cos\Theta_s)\right]\, .
\nonumber \\
\een
The small $\Theta_s$ expansion of this function takes the form:
\ben\label{eb17}
\int  {d E\over d\omega d\Omega} d\Omega &=&
{E^2\over \pi^2} \, {\Theta_s^2\over 2} \left\{1 + 2\ln 2 +\ln\Theta_s^{-2} +\OO(\Theta_s^3)\right\}
 \\ &+& {E^4\over 2\pi^4} \omega^2 (\ln\omega)^2 \left[1 -  {\Theta_s^2\over 2} \left\{1 + 2\ln 2 +\ln\Theta_s^{-2}\right\}+\OO(\Theta_s^3)\right] +\OO(\omega^2\ln\omega)\, .\nonumber
\een

Even though we have evaluated \refb{eb16} for general $\Theta_s$, during this derivation we have
ignored the possible modification of $A_{\mu\nu}$ due to radiation emitted during the 
scattering. Now for scattering at large impact parameter $b$, we have 
$\Theta_s\sim E/b$\cite{amati}. If we assume that
the spectrum of gravitational radiation  falls off rapidly for 
$\omega>b^{-1}\sim\Theta_s/E$, then
integrating \refb{eb17} in the range $0\le\omega\le \Theta_s/E$ we see that the total
radiated energy during the scattering is of order $E\Theta_s^3$ times possible factors
of $\ln\Theta_s^{-1}$. Since the correction to $\eps^\pm_{\mu\nu}A^{\mu\nu}$ 
from a final state particle is proportional
to the energy carried by the particle, we see that $\eps^\pm_{\mu\nu}A^{\mu\nu}$ 
can receive correction of
order $E\Theta_s^3$, and this in turn can affect the coefficients appearing in
\refb{eb16} by terms of order $\Theta_s^3$. However expansion up to order $\Theta_s^2$,
given in \refb{eb17} can be trusted.

It was shown in \cite{1409.4555} however that in the near forward direction, the actual cut-off on
$\omega$ extends beyond $1/b$ and as a result the net energy of emitted radiation is of 
order $\Theta_s^2$ with possible logarithmic corrections. Therefore one might worry that this will
give corrections to $\eps^\pm_{\mu\nu}A^{\mu\nu}$ of order $\Theta_s^2$ and affect the 
order $\Theta_s^2$ coefficient of the $\omega^2(\ln\omega)^2$ term. However one can see as
follows that this is not the case. If the energy is emitted in the near forward direction, then it
effectively amounts to a redistribution of the energy among the various final state
particles in the
forward direction and does not affect $A_{\mu\nu}$. For example a final state particle of
momentum $\lambda p$ and another particle of momentum $(1-\lambda)p$ give the same
contribution to $A_{\mu\nu}$ as a single particle of momentum $p$. Using this one finds that
for finite $\theta$, the correction to $\eps^\pm_{\mu\nu}A^{\mu\nu}$ is still of order $\Theta_s^3$
(possibly multiplied by powers of $\ln\Theta_s^{-1}$)
while for $\theta\sim\Theta_s$ the correction to $\eps^\pm_{\mu\nu}A^{\mu\nu}$ is of
order $\Theta_s^2$. This in turn can be used to show that the correction to \refb{eb16} due to the
modification of $A_{\mu\nu}$ by the final state radiation is of order $\Theta_s^3$ times possible
logarithmic corrections.

The positivity of the coefficient of the $\omega^2(\ln\omega)^2$ term for small $\Theta_s$
shows that the flux has a local minimum at $\omega=0$ and therefore has a maximum elsewhere,
presumably around $\omega\sim b^{-1}$. This confirms the prediction of 
\cite{1812.08137}. The actual coefficient of this term differs from that of \cite{1812.08137} by
a factor of 2 in the small $\Theta_s$ limit. On the other hand the coefficient of the 
$E^2\Theta_s^2/(2\pi^2)$ term differs from that of \cite{1812.08137} by the additive constant
$2\ln 2$. Both of these can be attributed to the fact that \cite{1812.08137} computed the
differential flux in the small $\theta$ approximation, whereas \refb{eb17} receives contribution
also from the finite $\theta$ region.

\bigskip

\noindent {\bf Acknowledgement:} 
We wish to thank Dimitri Colferai and Gabriele Veneziano for useful discussions, 
comments on an earlier version of
the draft and encouraging us to do the detailed analysis described in \S\ref{sb}.
B.S. is supported by the Simons Foundation grant 488649 (Simons
Collaboration on the Nonperturbative Bootstrap) and by the Swiss
National Science Foundation through the National Centre of Competence
in Research SwissMAP.
The work of A.S. was
supported by the  Infosys chair professorship and the
J. C. Bose fellowship of 
the Department of Science and Technology, India. 

\appendix

\sectiono{Effect of electromagnetic interaction} \label{sa}

If the incoming and the outgoing particles carry electric charge then the coefficients
$B_{\mu\nu}$, $C_{\mu\nu}$, $F_{\mu\nu}$ and $G_{\mu\nu}$ receive additional 
corrections\cite{1808.03288,1912.06413,2008.04376}. In this appendix we shall show
that even in the presence of these corrections, $B_{\mu\nu}$ and $F_{\mu\nu}$
continue to be independent of the momenta (and charges) 
carried by individual massless particles in the
final state.

The extra contribution to $B_{\mu\nu}$ due to electromagnetic interaction is given 
by\cite{1912.06413}:
\be \label{ea.1}
\Delta B^{\mu\nu} = -\, {2\, G\over R\, c^5} \left[
\sum_{i=1}^n \sum_{j=1\atop j\ne i}^n 
{1\over 
\{(p_{i}.p_{j})^2 
-p_{i}^2 p_{j}^2 \}^{3/2}} \, 
 \right. 
\left. 
 \, {p_{i}^\mu \over n.p_{i}}\,
 (n.p_{j}\, p_{i}^\nu - n.p_{i}\, p_{j}^\nu ) \times \frac{1}{4\pi}q_i q_j p_i^2 p_j^2
 \right]\, ,  
 \ee
where $q_i$ is the charge carrried by the $i$-th final state particles, in units where the
electrostatic force between  a pair of charges separated by distance $r$ 
is given by $q_iq_j/ (4\pi r^2)$. 
\refb{ea.1} 
clearly vanishes if either $i$ or $j$ represents a massless particle. 

The correction to $F^{\mu\nu}$ is given by\cite{2008.04376}:
\ben \label{ea.2}
&&  \Delta F^{\mu\nu} \non\\
&  =&   {2\, G\over R\, c^{9}}\, \Bigg[  
 -4\, G\, \sum_{\ell=1}^n p_{\ell}.n \sum_{i=1}^{n} \sum_{j=1\atop j\ne i}^{n}
 {1\over p_{i}.n} {1\over \{(p_{i}.p_j)^2 - p_{i}^2 p_j^2\}^{3/2}} 
  \{ n.p_j \, p_{i}^\mu \, p_{i}^\nu - n.p_{i} \, p_{i}^\mu \, p_j^\nu\}\times \frac{q_i q_j}{4\pi}p_i^2 p_j^2
 \non\\ &&   
-2\, G\, \sum_{\ell=1}^m p'_{\ell}.n \sum_{i=1}^{m} \sum_{j=1\atop j\ne i}^{m}
 {1\over p'_{i}.n} {1\over \{(p'_{i}.p'_j)^2 - p_{i}^{\prime 2} p_j^{\prime 2}\}^{3/2}} 
  \{ n.p'_j \, p_{i}^{\prime\mu} \, p_{i}^{\prime \nu} - 
 n.p'_{i} \, p_{i}^{\prime \mu} \, p_j^{\prime \nu}\}  \times \frac{q'_i q'_j}{4\pi}p_i^{\prime 2} p_j^{\prime 2}\non\\ &&   
-2\, G  \sum_{i=1}^{n} \sum_{j=1\atop j\ne i}^{n}
\sum_{\ell=1\atop \ell\ne i}^{n}  {1\over p_{i}.n}  
{p_{i}.p_j\over \{(p_{i}.p_j)^2 - p_{i}^2 p_j^2\}^{3/2}}   
\{2 (p_{i}.p_j)^2 - 3 p_{i}^2 p_j^2\} {1\over \{(p_{i}.p_{\ell})^2 - p_{i}^2 p_{\ell}^2\}^{3/2}} \non\\ &&
\times \frac{q_i q_\ell}{4\pi} p_i^2 p_{\ell}^2\,  \{ n.p_j \, p_{i}^\mu  - n.p_{i} \, p_j^\mu \} \, 
\{ n.p_{\ell} \, p_{i}^\nu - n.p_{i} \, p_{\ell}^\nu \}\nonumber\\
&& 
+ \ c^2\, \sum_{i=1}^{n} \sum_{j=1\atop j\ne i}^{n}
\sum_{\ell=1\atop \ell\ne i}^{n}  {1\over p_{i}.n}  
{1\over \{(p_{i}.p_j)^2 - p_{i}^2 p_j^2\}^{3/2}}   
 {1\over \{(p_{i}.p_{\ell})^2 - p_{i}^2 p_{\ell}^2\}^{3/2}}\, \times \frac{q_i q_j}{4\pi} p_i^2 p_{j}^2\, \times \frac{q_i q_\ell}{4\pi} p_i^2 p_{\ell}^2\,  \non\\ &&
 \{ n.p_j \, p_{i}^\mu  - n.p_{i} \, p_j^\mu \} \, 
\{ n.p_{\ell} \, p_{i}^\nu - n.p_{i} \, p_{\ell}^\nu \}
\Bigg]\, .
\een
It is understood that the expression needs to be symmetrized under the exchange of $\mu$ and $\nu$.
As before we divide the final state particles into massive particles carrying momenta $\wt p_i$ and
charges $\wt q_i$ and massless particles with momenta $\wh p_i$ and charges $\wh q_i$.
Examining this expression we see that the only contributions from massless final state particles
can come when $p_\ell$ represents a massless particle momentum in the first term inside the
square bracket and when $p_j$ represents a massless particle momentum in the third term
inside the square bracket. The first contribution may be expressed as:
\be\label{ea.3}
-4\, G\, \sum_\ell  \wh p_{\ell}.n \sum_i \sum_{j\atop j\ne i}
 {1\over \wt p_{i}.n} {1\over \{(\wt p_{i}.\wt p_j)^2 - \wt p_{i}^2 \wt p_j^2\}^{3/2}} 
  \{ n.\wt p_j \, \wt p_{i}^\mu \, \wt p_{i}^\nu - n.\wt p_{i} \, \wt p_{i}^\mu \, \wt p_j^\nu\}\times \frac{\wt
  q_i \wt q_j}{4\pi}\wt p_i^2 \wt p_j^2\, . \label{step}
\ee
On the other hand when the sum over $j$ runs over massless particles in the third term 
within the square bracket, the contribution takes the form:
\be
4\, G  \sum_j \sum_i
\sum_{\ell\atop \ell\ne i} {1\over \wt p_{i}.n}  {1\over \{(\wt p_{i}.\wt p_{\ell})^2 
- \wt p_{i}^2 \wt p_{\ell}^2\}^{3/2}} 
\times \frac{\wt q_i \wt q_\ell}{4\pi} \wt p_i^2 \wt p_{\ell}^2\,  \{ n.\wh p_j \, 
\wt p_{i}^\mu  - n.\wt p_{i} \, \wh p_j^\mu \} \, 
\{ n.\wt p_{\ell} \, \wt p_{i}^\nu - n.\wt p_{i} \, \wt
p_{\ell}^\nu \}\, .
\ee
In the above expression, relabelling $\ell$ as $j$ and $j$ as $\ell$ 
and simplifying we get:
\ben\label{ea.5}
&&4\, G   \sum_\ell \sum_i
\sum_{j\atop j \ne i}  {1\over \wt p_{i}.n}  {1\over \{(\wt p_{i}.\wt p_{j})^2 - \wt p_{i}^2 \wt p_{j}^2\}^{3/2}} 
\times \frac{\wt q_i \wt q_j}{4\pi} \wt p_i^2 \wt p_{j}^2\,  \{ n.\wh p_\ell \, \wt p_{i}^\mu  - n.\wt p_{i} \, 
\wh p_\ell^\mu \} \, 
\{ n.\wt p_{j} \, \wt p_{i}^\nu - n.\wt p_{i} \, \wt p_{j}^\nu \}\nonumber\\
&=& 4\, G\, \sum_{\ell} \wh p_{\ell}.n \sum_{i} \sum_{j\atop j\ne i}
 {1\over \wt p_{i}.n} {1\over \{(\wt p_{i}.\wt p_j)^2 - \wt p_{i}^2 \wt p_j^2\}^{3/2}} 
  \{ n.\wt p_j \, \wt p_{i}^\mu \, \wt p_{i}^\nu - n.\wt p_{i} \, \wt p_{i}^\mu \, \wt p_j^\nu\}\times \frac{\wt q_i \wt q_j}{4\pi}\wt p_i^2 \wt p_j^2 \nonumber\\
  && -4\, G\, \sum_{\ell} \sum_{i} \sum_{j\atop j\ne i}
  {1\over \{(\wt p_{i}.\wt p_j)^2 - \wt p_{i}^2 \wt p_j^2\}^{3/2}} 
   \wh p_\ell^\mu \{ n.\wt p_j \, \wt p_{i}^\nu - n.\wt p_{i}  \, \wt p_j^\nu\}\times \frac{\wt q_i \wt q_j}{4\pi}\wt p_i^2 \wt p_j^2\, .
\een
We now see that the first term on the right hand side of \refb{ea.5}
cancels \eqref{step} and the second term vanishes since the summand is anti-symmetric 
under the exchange of  $i$ and $j$. 
Hence we do not 
get any contribution involving final state massless particles due to long-range
electromagnetic interaction.

\end{document}